\documentclass[twocolumn]{aastex62}

\usepackage[fleqn]{amsmath} 
\usepackage{amssymb}  
\usepackage{gensymb}
\usepackage[nolist]{acronym}
\usepackage{ulem}
\usepackage{bm}
\usepackage{etoolbox}
\usepackage{hyperref}
\usepackage{enumitem}

\definecolor{ochre}{rgb}{0.8, 0.47, 0.13}

\newcommand{\customlabel}[2]{%
   \protected@write \@auxout {}{\string \newlabel {#1}{{#2}{\thepage}{#2}{#1}{}} }%
   \hypertarget{#1}{}
}

\usepackage{import}

\usepackage{xspace}
\shorttitle{Formation of heavy BNSs and light BH-NSs}
\shortauthors{Vigna-G\'omez et al.}

\begin{document}

\title{Fallback supernova assembly of heavy binary neutron stars and light black hole$-$neutron star pairs and the common stellar ancestry of GW190425 and GW200115}

\correspondingauthor{Alejandro Vigna-G\'omez}
\email{avignagomez@nbi.ku.dk}

\author[0000-0003-1817-3586]{Alejandro Vigna-G\'omez}
\affil{DARK, Niels Bohr Institute,
University of Copenhagen,
Jagtvej 128, 2200,
Copenhagen, Denmark}
\affil{Niels Bohr International Academy, The Niels Bohr Institute, Blegdamsvej 17, 2100 Copenhagen, Denmark.}

\author[0000-0003-1735-8263]{Sophie L. Schr\o der}
\affil{DARK, Niels Bohr Institute,
University of Copenhagen,
Jagtvej 128, 2200,
Copenhagen, Denmark}

\author[0000-0003-2558-3102]{Enrico Ramirez-Ruiz}
\affil{Department of Astronomy and Astrophysics,
University of California,
Santa Cruz, CA 95064, USA}
\affil{DARK, Niels Bohr Institute,
University of Copenhagen,
Jagtvej 128, 2200,
Copenhagen, Denmark}

\author[0000-0002-3874-2769]{David R. Aguilera-Dena}
\affil{Institute of Astrophysics, FORTH, 
Dept. of Physics, University of Crete, 
Voutes, University Campus, 
GR-71003 Heraklion, Greece}
\affil{Argelander-Institut f\"ur Astronomie, 
Universit\"at Bonn, 
Auf dem H\"ugel 71, 
53121 Bonn, Germany}
\affil{Max-Planck-Institut f\"ur Radioastronomie, 
Auf dem H\"ugel 69, 
53121 Bonn, Germany}

\author[0000-0002-3269-3847]{Aldo Batta}
\affil{Instituto Nacional de Astrofísica, \'Optica y Electr\'onica, 
Tonantzintla, 
Puebla 72840, M\'exico}

\author{Norbert Langer}
\affil{Argelander-Institut f\"ur Astronomie, 
Universit\"at Bonn, Auf dem H\"ugel 71,
53121 Bonn, Germany}
\affil{Max-Planck-Institut f\"ur Radioastronomie, 
Auf dem H\"ugel 69, 
53121 Bonn, Germany}

\author[0000-0003-0674-9453]{Reinhold Willcox}
\affil{School of Physics and Astronomy, 
Monash University, Clayton, 
Vic. 3800, Australia}
\affil{The ARC Centre of Excellence for Gravitational Wave Discovery -- OzGrav}



\begin{abstract}
The detection of the unusually heavy binary neutron star merger GW190425 marked a stark contrast to the mass distribution from known Galactic pulsars in double neutron star binaries and gravitational-wave source GW170817.
We suggest here a formation channel for heavy binary neutron stars and light black hole - neutron star binaries in which massive helium stars, which had their hydrogen envelope removed during a common envelope phase, remain compact and avoid mass transfer onto the neutron star companion, possibly avoiding pulsar recycling.
We present three-dimensional simulations of the supernova explosion of the massive stripped helium star and follow the mass fallback evolution and the subsequent accretion onto the neutron star companion. 
We find that fallback leads to significant mass growth in the newly formed neutron star.
This can explain the formation of heavy binary neutron star systems such as GW190425, as well as predict the assembly of light black hole - neutron star systems such as GW200115.
This formation avenue is consistent  with the observed mass-eccentricity correlation of binary neutron stars in the Milky Way.
Finally, avoiding mass transfer suggests an unusually long spin-period population of pulsar binaries in our Galaxy.

\end{abstract}




\section{Introduction}
\label{sec:intro}
On April 25th, 2019, the LIGO-Virgo network detected its second-ever signal of two neutron stars merging, tagged as GW190425 \citep{abbott2020gw190425}. 
But unlike the first detection of a binary neutron star (BNS) merger \citep[GW170817,][]{GW170817}, which conformed to expectations, GW190425 was extraordinary. 
Most of what we know about neutron stars comes primarily from observations of  pulsars, magnetized rotating neutron stars, in our own Milky Way. 
Of the thousands of known pulsars, almost twenty are visible as recycled millisecond pulsars paired with another neutron star companion \citep{Tauris2017formation,AndrewsMandel2019}. These light neutron star binaries, including GW170817, weighed the equivalent of about 2.6 solar masses \citep{Kiziltan2013,ozel2016masses,Farrow2019ApJ}. 
By contrast, GW190425 has a total mass equal to about 3.4 solar masses \citep{abbott2020gw190425}. 

Since the detection of the Hulse–Taylor binary \citep{HulseTaylor1975}, there is consensus that the progenitors of BNSs are massive stellar binaries \citep[e.g.,][]{Heuvel1976closeBinaries}.
A crucial phase in the evolutionary pathway to BNS formation occurs when a giant star fills its Roche lobe and initiates a dynamically-unstable mass-transfer episode onto the neutron star companion  \citep[e.g.,][]{bhattacharya1991formation}.
The stellar core and the neutron star become engulfed by the expanding envelope, a process where gas drag dissipates orbital energy of the binary \citep[e.g.,][]{Ivanova2013,2015ApJ...798L..19M}.
This common envelope phase ends when the hydrogen envelope is ejected and a compact, stripped, helium-rich star of a few solar masses  is left to reside  in a tight ($\approx \rm R_{\odot}$) near circular orbit  \citep{Fragos2019,LawSmith2020}.
The subsequent evolution of the binary (Figure 1) depends on the mass and composition of the stripped helium star \citep{Woosley2019} after the envelope is ejected.
Most low-mass helium stars expand \citep[e.g.,][]{Woosley1995,Gotberg2017,2020A&A...637A...6L} and engage in an additional stable mass-transfer episode.
During this episode, the mass transferred from the helium-rich donor recycles the pulsar, a process in which the neutron star spin increases to milliseconds and becomes radio visible for several Gyr \citep[e.g.,][]{SRINIVASAN201093}.
Moreover, the donor star becomes an ultra-stripped core \citep{Tauris2013,Tauris2015ultra}.
These low-mass systems lead to BNSs such as GW170817 and those observed in the Milky Way \citep[e.g.,][]{2019ApJ...883L...6R}.

\begin{figure}
    \centering
    \includegraphics[width=0.47\textwidth]{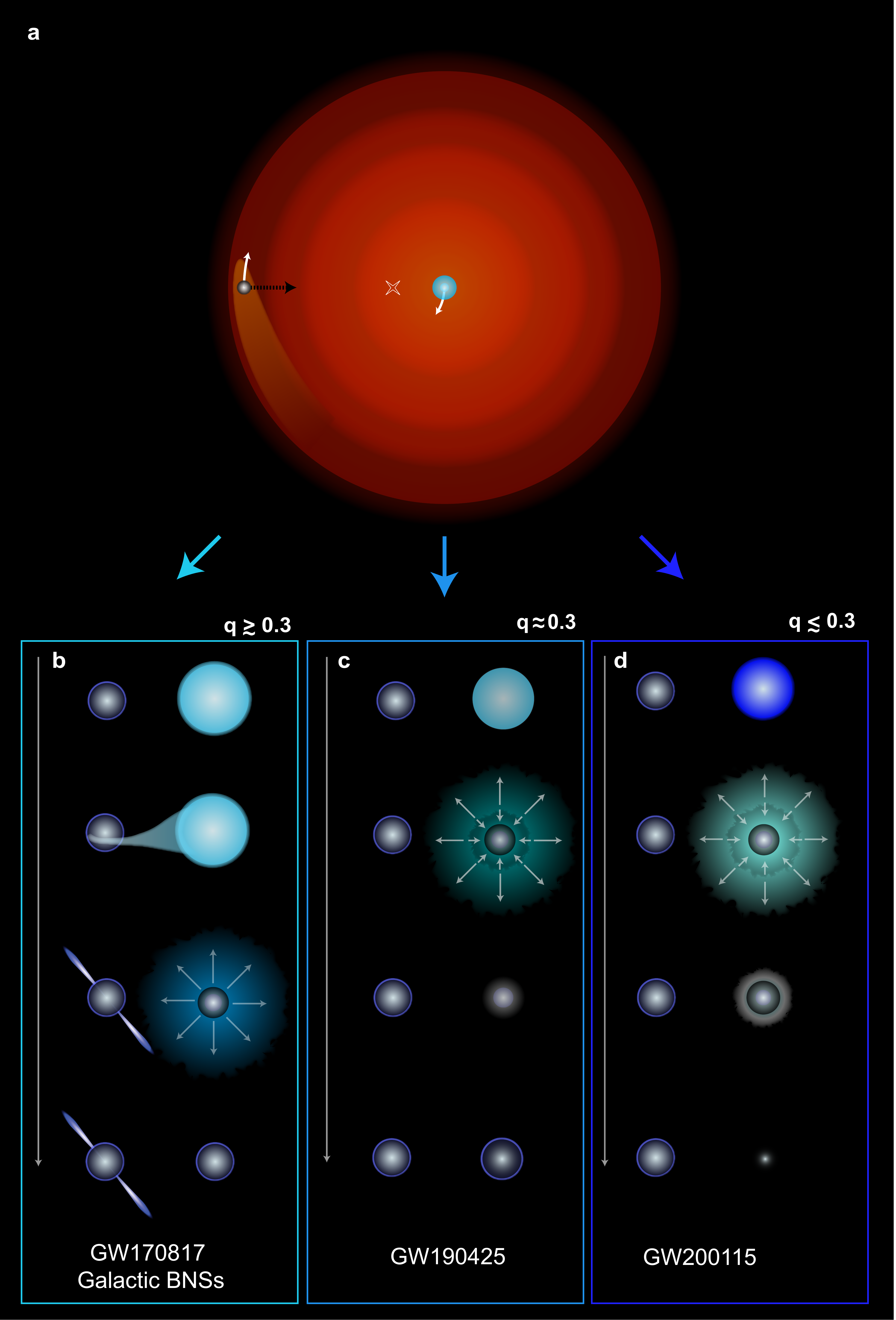}  
    \caption{
    Late stages of BNS formation.
    The giant star expands and engulfs the neutron star companion in an stage commonly referred to as a common-envelope evolution (\textbf{a}).
    A successful ejection of the envelope leaves the neutron star in a close orbit with a stripped-envelope star.
    The evolution of the system depends on the mass ratio $q=M_{\rm{NS}}/M_{\rm{stripped}}$.
    Less-massive stripped stars with $q\gtrsim 0.3$ experience an additional mass transfer phase that further strips the star and recycles the pulsar companion. Such evolutionary sequence  leads to systems such as the observed BNSs in the Milky Way and GW170817 (\textbf{b}).
    More massive stripped stars with $q\approx 0.3$ do not expand as much, therefore avoiding further stripping and companion recycling. Such evolutionary sequence, on the other hand, is expected to lead to BNS systems  such as GW190425 (\textbf{c}).
    Finally, even more massive stripped stars with $q\lesssim 0.3$ will lead to BH-NS binaries such as GW200115 (\textbf{d)}.
    }
    \label{fig:cartoon}
\end{figure}

In this \textit{Letter} we propose an alternative channel formation channel for heavy BNSs.
In this formation channel, massive ($\gtrsim 9\ \rm{M_{\odot}}$) helium stars remain compact and avoid mass transfer onto a neutron star and thus pulsar recycling.
Non-recycled, young, pulsars become radio quiet after only tens of Myr \citep[e.g.,][]{2004hpa..book.....L,Tauris2017formation} and, as a result,  these massive helium stars could lead to radio-quiet compact binaries that can only be detected by gravitational-wave observatories.
The structure of the helium star at core collapse will determine if the system will become a BNS or black hole - neutron star (BH-NS) binary.
These systems offer an alternative evolutionary pathway which can explain the dichotomy between the observed BNSs hosting recycled pulsars, GW170817, and the unusually heavy gravitational-wave source GW190425 (Figure \ref{fig:cartoon}).

\section{Methods and Initial Conditions}
\label{sec:method}
In this \textit{Letter} we present three-dimensional (3D) hydrodynamic models of GW190425-like progenitor binaries using the smoothed-particle hydrodynamics (SPH) code \textsc{GADGET-2} \citep{Springel2005Gadget}.
Our approach is hybrid, since we approximate the evolution of a star that is stripped by a binary companion through a detailed stellar evolution of a single star, then map the pre-collapse 1D stellar model onto a 3D binary to model the explosion.
To generate our initial models we make use of the one-dimensional (1D) stellar evolution code MESA \citep{Paxton2011MESA} version 10398. 
In particular,  we model the evolution of a $10.0\ \rm M_{\odot}$ stripped star at $Z=0.02$ from helium zero-age main sequence to core collapse. 
At core collapse, the heavy helium star has a mass of $M_{\rm{pre-SN}} \approx 5.4\ \rm M_{\odot}$, a radius of $\approx 0.7\ \rm{R_{\odot}}$, and more than 95\% of its gravitational binding energy contained below a radius of $0.01\ \rm R_{\odot}$.  The reader is refer to Appendix~\ref{app:MESA} for specifics. We then map the MESA model into \textsc{GADGET-2} in order to
simulate the supernova explosion of a heavy helium star with a 1.3 $\rm{M_{\odot}}$ neutron star companion at a separation of $a_{\rm{pre-SN}} = 1.4\ \rm{R_{\odot}}$ in a circular orbit. Details on the setup and numerical tests can be found in  Appendix~\ref{app:Gadget}.
The initial proto-neutron-star mass is assumed to be $M_{\rm{proto-NS}} = 1.3\ \rm{M_{\odot}}$, consistent with the observed mass distribution of BNSs \citep{Kiziltan2013,ozel2016masses,Farrow2019ApJ} and with the properties of the 1D pre-supernova stellar model \citep{Muller2016}.
An explosion energy of 1.5 bethes\footnote{$1\,\rm{bethe} := 10^{51}\ \rm{erg}$.}, consistent with estimates from a 1D neutrino-hydrodynamics code for a similar progenitor model \citep{Ertl2020}, is deposited in a shell above the proto-neutron star.
We focus on the long-term, post-explosion fallback evolution of the ejecta in order to account for mass accretion of the newly born neutron star and for pulsar recycling of the companion.
We do not account for magnetic fields.

\section{Neutron Star Birth from Supernova Fallback}
\label{sec:results}
\begin{figure}
    \centering
    \includegraphics[trim=0.5cm 0 0 0,clip,width=0.52\textwidth]{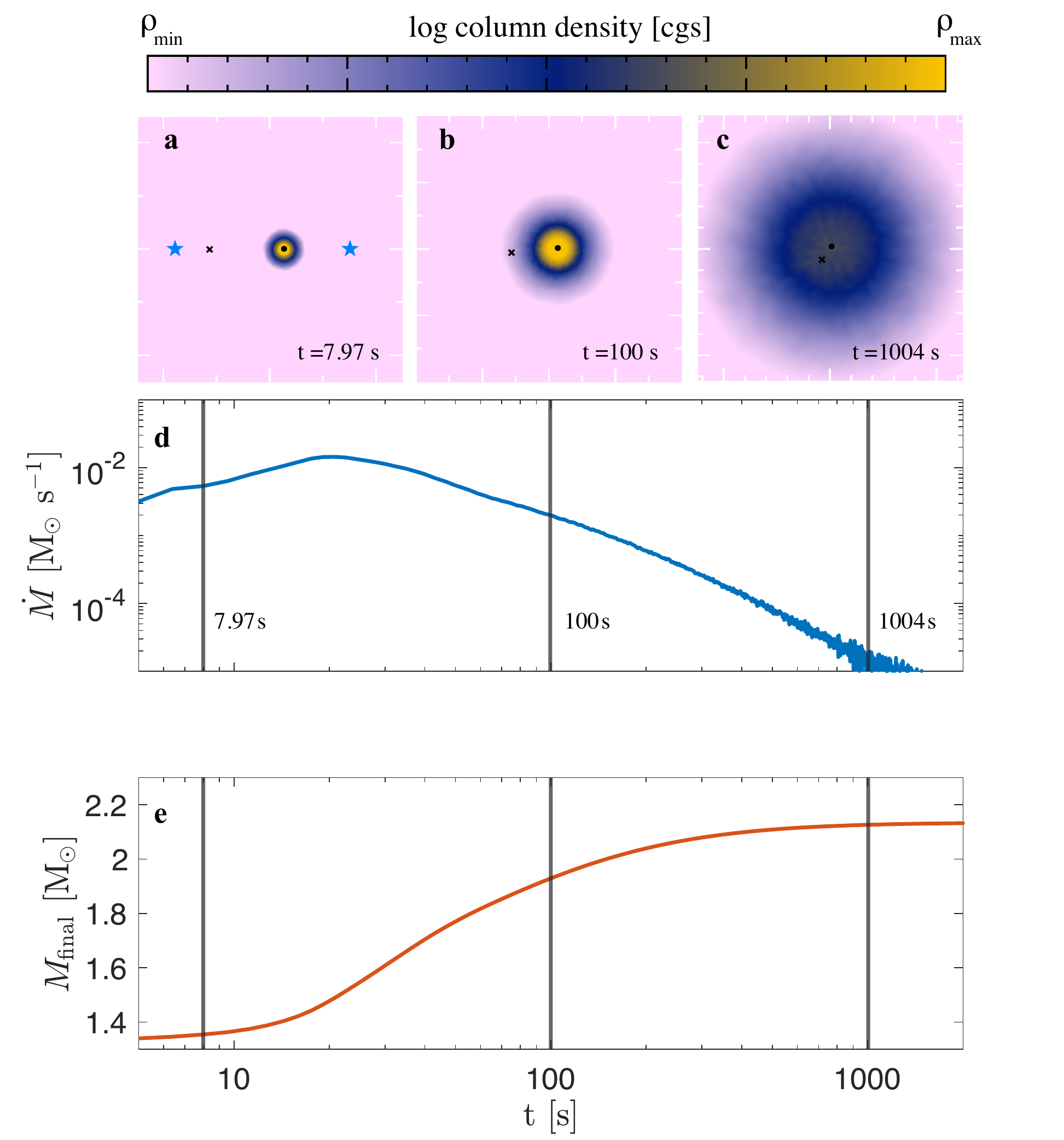}
    \caption{
    The hydrodynamical evolution of the second supernova and the accompanying mass fallback that leads to a heavy BNS merger.
    Panels \textbf{(a)-(c)} show the column density (cgs units) in base 10 logarithmic scale and span [-1,3] in \textbf{(a)}, [-2,2] in \textbf{(b)} and [-3,1] in \textbf{(c)}.
    The location of the newly born neutron star is shown as a filled black circle and the companion neutron star is shown as a black cross.    
    The second and third outer Lagrangian points of the binary are shown as blue stars in panel \textbf{(a)}.
    The tick marks on each panel correspond to a solar-radius scale.
    The only interaction with the neutron star companion is from the blasted ejecta and there is only a tiny mass of material accreted, implying that the pulsar companion will not be effectively recycled. 
    Panels \textbf{(d)} and \textbf{(e)} show the fallback mass accretion rate onto the newly born neutron star and its cumulative mass accretion growth, both with vertical lines marking the snapshots from panels \textbf{(a)-(c)}.
    }
    \label{fig:hydro}
\end{figure}

The resultant hydrodynamical evolution of the explosion is depicted in Figure 2. 
The shock initially propagates through the iron core until it reaches the envelope, fractions of a second after the explosion.
At this point, a reverse shock wave emerges, which  propagates back towards the newly-formed neutron star and triggers   mass fallback. 
The fallback mass accretion rate peaks 20 s after the explosion at $\approx 10^{-2}\ \rm M_{\odot}\ s^{-1}$ (Figure \ref{fig:hydro}).
Approximately $0.8\ \rm{M_{\odot}}$ are accreted during the first hundred seconds after the explosion, roughly the same time scale in which the expanding layers of the exploding star reach the neutron star companion (Figure \ref{fig:hydro}).
The rapid velocity of the expanding shock ($\rm \approx 1000\ km\ s^{-1})$ and the small cross section of the neutron star companion result in $\lesssim 10^{-3}\ \rm{M_{\odot}}$ of accreted material.
The accretion of this small amount of material will not effectively recycle the neutron star companion \citep[e.g.,][]{Tauris2017formation}.

After thousands of seconds the newly formed  neutron star approaches a final mass of $\approx 2.1\ \rm{M_{\odot}}$, a value in broad agreement with earlier results \citep{fryer2012compact,Ertl2020} in the literature.
During the whole simulation the accretion rate remains above hypercritical \citep{Chevalier1993ApJ} and  neutrinos provide the main cooling mechanism until after $\approx 10^6$ s. 

\begin{figure}
    \centering
    \includegraphics[trim=1.5cm 7.0cm 2cm 7.0cm,clip,width=\columnwidth]{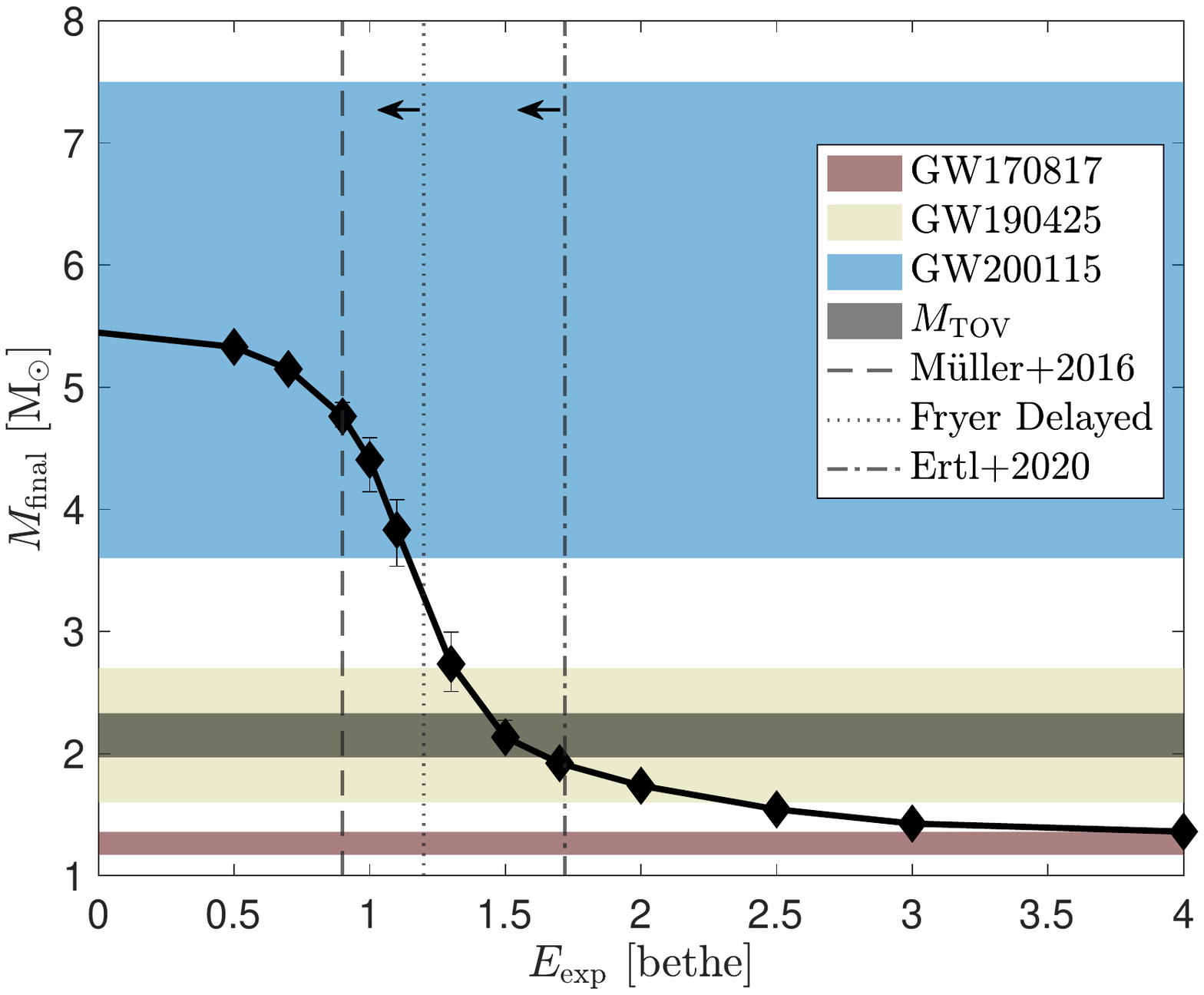}
    \caption{
    Mass of the newly formed remnant as a function of the supernova explosion energy after the fallback accretion has ceased.
    The black diamonds represent the models explored in this work, and the thick solid black line is a linear interpolation between the values.
    The range of the component mass for gravitational-wave sources GW170817, GW190425, and GW200115 is shown in pink, beige, and blue, respectively, while the maximum mass of a non-rotating neutron star, $M_{\rm{TOV}}$ \citep{Rezzolla2018}, in shown in gray.
    For reference, the birth mass of the neutron star is $1.3\ \rm{M_{\odot}}$, same as the mass of the neutron star companion.
    The semi-analytical prediction \citep{Muller2016} of the explosion energy for this particular model is shown as a dashed black line.
    The upper limits of supernova models with explosion energies which include fallback are shown as dotted and dash-dotted lines \citep{fryer2012compact,Ertl2020}.
    Main numerical uncertainties are included as error bars, some of them within the symbols (Appendix \ref{app:Gadget}).
    }
    \label{fig:explosions}
\end{figure}

The amount of fallback mass accretion  increases with decreasing  explosion energy (Figure \ref{fig:explosions}).
Energies of $E_{\rm{exp}} \lesssim 0.5$ bethes lead to almost complete fallback while explosions energies of $E_{\rm{exp}} \gtrsim 2.5$ bethes lead to a complete ejection of the envelope.
The fallback-dominated transition from neutron star to black hole remnants occurs at  $E_{\rm{exp}} \lesssim 1.3$ bethes.
Explosion energies between $1.3 \lesssim E_{\rm{exp}} \lesssim 2.4$ bethes lead to remnant masses $1.6 \lesssim M_{\rm{rem,exp}}/\rm{M_{\odot}} \lesssim 2.7$, which are in the inferred range for the heavy  neutron star in GW190425  \citep{abbott2020gw190425}.
Future detections of BNSs and BH-NS binaries would thus help improve the so far weak constrains of supernova explosion energies from massive helium stars.

The ejected envelope material during a supernova explosion imparts a recoil kick on the system.
Even if  the supernova is spherically-symmetric in the frame of reference of the exploding star, the explosion will increase the orbital period and eccentricity \citep{Blaauw1961}. 
If, on the other hand, the supernova material is ejected anisotropically, the magnitude of the resultant kick to the newly born neutron star is expected to be of the order of $\approx 100\ \rm{km\ s^{-1}}$ for isolated massive stars \citep[e.g.,][]{BurrowsVartanyan2021Nature} and reduced to $\approx 10\ \rm{km\ s^{-1}}$ for ultra-stripped or electron-capture supernovae \citep[e.g.,][]{VignaGomez2018}.
BNSs assembled via common-envelope episodes end up in close orbits with relative orbital velocities well in excess of $1000\ \rm{km\ s^{-1}}$ and are likely to remain gravitationally bound after the explosion.
Depending on the direction and  magnitude of the natal kick, some binaries might actually end up shrinking to even closer orbits.
The explosion of massive helium stars with a light neutron star companion are expected to lead to the formation of more eccentric binaries. 

\section{Discussion and Conclusions} 
\label{sec:discAndConc}

\subsection{Light BH-NS binaries and GW200115}
\label{subsec:NSBHs}
On January 15th, 2020, the LIGO-Virgo network detected GW200115, the second ever confident detection of a BH-NS coalescence \citep{2021ApJ...915L...5A}.
GW200115 is composed of a neutron star and a black hole with masses of $1.5^{+0.7}_{-0.3}\ \rm{M_{\odot}}$ and $5.7^{+1.8}_{-2.1}\ \rm{M_{\odot}}$, respectively.
Such light BH-NS can be assembled via isolated binary evolution according to population studies \citep{2021arXiv210805763B}.
However, the component masses of GW200115 are peculiar.
The mass of the neutron star is marginally more massive the $1.33\ \rm{M_{\odot}}$ mean observed in Galactic BNSs mass distribution \citep{Farrow2019ApJ}.
The black hole is close to lower side of the black hole mass distribution.
This mass can be easily explained by the low explosion energies from our model (Figure \ref{fig:explosions}).

According to the fallback model presented in this \textit{Letter}, a helium star of $10.0\ \rm M_{\odot}$ forms either a heavy neutron star or a light black hole.
However, semi-analytical and numerical models predict that the remnant mass function does not necessarily increases monotonically with the mass at core collapse \citep{2016ApJ...821...38S,Muller2016,Ertl2020}, and that the outcome depends on the structure of the stellar model as well as stochasticity in the explosion mechanism.
Therefore, similar stars in similar binaries could lead to both heavy BNSs and light BH-NSs, and the fallback explosion mechanism might explain both simultaneously.
The absence of one population could serve as a constrain on the explosion energy of stripped stars.

\subsection{Mass-eccentricity correlation}
There are hints of a mass-eccentricity correlation in short period ($<$ 1 day) BNSs in the Milky Way, where millisecond pulsars paired with more massive companions ($\approx 1.4\ \rm{M_{\odot}}$) are in more eccentric ($\approx 0.6$) orbits \citep[e.g.,][]{Tauris2017formation,AndrewsMandel2019}.
The formation channel proposed here for GW190425 is consistent with this trend, as mass loss during the second supernova in heavy BNS formation can lead to large eccentricities.
In contrast, the formation of light BH-NSs such as GW200115 will result in decreased mass loss during the second supernova, and would lead to low eccentricities at double compact object formation.
The fallback scenario presented here thus provides an  explanation for the observed mass-eccentricity correlation without the need to rely on a dynamical-formation scenario \citep{AndrewsMandel2019}.
To date, there is no evidence of heavy ($> 2.9\ \rm{M_{\odot}}$) BNSs in the Milky Way.
This suggests at least one of the following three things about heavy BNSs: they have very short orbital periods ($\lesssim$ few hours) and thus avoid detection in acceleration searches \citep{abbott2020gw190425,Safarzadeh2020GW190425,Galaudage2021}, they are radio quiescent, or such systems are rare in the Milky Way \citep{2020A&A...639A.123K,Galaudage2021}.
A priori, there is no reason why heavy BNSs should be preferentially born in short orbital periods \citep[but see][]{2020MNRAS.496L..64R} and standard formation models are unable to predict enough fast mergers to be reconciled with the detection of GW190425 \citep{Safarzadeh2020GW190425}.

\subsection{Electromagnetic counterparts and gravitational waves}
The merger of a heavy neutron star pair or a light BH-NS binary is expected to produce an electromagnetic counterpart that will further shed light on its origin \citep{Roberts2011ApJL}.
Particularly, the merger of a heavy neutron star pair is expected to produce a luminous red kilonova likely powered by an accretion disc wind \citep{2017Natur.551...80K}, which might likely be accompanied by a blue kilonova component \citep{2014MNRAS.441.3444M}. 
The merger of a light BH-NS binary, on the other hand, is expected to experience tidal disruption and  be only observable as faint red kilonova \citep{2017Natur.551...80K}. The accompanying electromagnetic signatures would provide a natural test to distinguish between different compact binary mergers.

The formation channel presented here hints to the presence of heavy BNSs or light BH-NSs in the Milky Way.
These and similar systems, such as non-recycled light BNSs \citep{BelczynskiKalogera2001}, are expected to be uncovered by the Laser Interferometer Space Antenna \citep{AmaroSeoane2017,Lau2020LISADNSs}.

\subsection{Some open questions in stellar binary evolution}
The evolution from zero-age main sequence to double compact formation is rather complex. 
We have assumed here that the evolution of the system follows the canonical assembly of BNSs \citep[e.g.,][]{bhattacharya1991formation,Tauris2017formation}, which includes a common-envelope phase of a giant star with a neutron star companion \citep{Fragos2019,LawSmith2020}.
\cite{VignaGomez2020} predicts that, at the onset of the common-envelope phase, only $\lesssim\ 5\%$ of neutron star binary progenitors will have donor stars with masses $\gtrsim\ 20\ \rm{M_{\odot}}$.
However, that study does not incorporate the recently explored stellar evolution models of stripped stars (Appendix~\ref{app:MESA}) nor the explosion mechanism explored  in this \textit{Letter} (Appendix~\ref{app:Gadget}).
These updates are likely to alter the predictions of assembly and merger rates for heavy BNSs and light BH-NS binaries.

Single unperturbed stellar models have been used to explore envelope ejection in massive binaries \citep{2016A&A...596A..58K,2021A&A...645A..54K}. 
Heavy ($\gtrsim\ 25\ \rm{M_{\odot}}$) progenitors with low-mass ($1\ \rm{M_{\odot}}$) companions are not likely to eject the envelope at high ($\approx$ solar) metallicities, a scenario which has been predicted to result in Thorne–{\. Z}ytkow objects \citep{1975ApJ...199L..19T,1977ApJ...212..832T}.
However, it is possible that modeling of progenitors with more massive companions \citep[cf. Figure 6 of][]{2021A&A...645A..54K}, lower metallicities, or different assumptions about energy requirements \citep{2020ApJ...899...77E,2021arXiv210714526V}, might lead to a successful ejection.

For models considered in this \textit{Letter}, we assumed that the orbit remains effectively unchanged after the envelope ejection.
However, the evolution of the post-common-envelope binary can entail energy-momentum transfer and losses via, e.g., stellar winds and tidal dissipation.
Mass loss via isotropic winds, aka the \textit{Jeans} mode, can widen the orbit by a factor of $\lesssim 2$, and in some cases counteract stellar expansion and therefore Roche-lobe overflow.
This is particularly relevant for stars with mass $6 \lesssim M_{\rm{stripped}}/\rm{M_{\odot}} \lesssim 10$ that will expand by a factor of a few at most.
Avoiding Roche-lobe overflow after the common-envelope phase would result in avoiding further pulsar recycling of the companion and leading to remnant masses $M_{\rm{rem,exp}} > 1.33\ \rm{M_{\odot}}$.
However, alternative mass loss modes or wind interaction with the companion could decrease the widening of the orbit \citep[e.g.,][]{2021arXiv210709675S}.

Throughout this \textit{Letter,} we do not consider tidal dissipation. 
The dynamical tide is unlikely to play a dominant role in the orbital evolution during late stages of BNS assembly, but it might (partially) counteract the widening of stellar winds.

\subsection{Mass accretion onto a neutron star and pulsar recycling}
A pulsar binary can be spun-up and recycled if angular momentum is efficiently transferred onto the pulsar.
This is a complex process that depends on the mass transfer rate, orbital properties of the binary, and accretion physics, as well as on the equation-of-state, magnetic field, and overall properties of the neutron star \citep{2012MNRAS.425.1601T,Tauris2017formation}.
Roche-lobe overflow from a stripped-star with a helium-rich envelope is an efficient way to form an accretion disc around the pulsar that can spin it up to tens of milliseconds and (mildy) recycle it; this is believed to be the preferred spin-up mechanism for Galactic BNSs \citep{Tauris2017formation}.
Avoiding such mass transfer episode, like we suggest in this \textit{Letter}, will avoid the main mass and angular momentum transfer mechanism, which for Galactic-like BNSs results in a mass growth of $6-9\times 10^{-3}\ \rm{M_{\odot}}$ and observed\footnote{We do not consider Galactic BNSs in Globular Clusters.} recycled pulsar spins between $17 < P_{\rm{spin}} < 186$ milliseconds \citep{Tauris2017formation,2018ApJ...854L..22S}.

However, post-common-envelope winds can also lead to mass accretion.
For the system presented in this \textit{Letter}, the amount of accreted mass $\Delta M_{\rm{acc}} = f_{\rm{acc}}\times \Delta M_{\rm{winds}}$, with $f_{\rm{acc}} \approx 10^{-4}$ is the estimated wind accretion efficiency \citep{Tauris2017formation} and $ \Delta M_{\rm{winds}} \approx 4.6\ \rm{M_{\odot}}$ is the amount of mass lost via stellar winds from the helium zero-age main sequence until core collapse (Appendix \ref{app:MESA}), results in $\Delta M_{\rm{acc}}\approx 4.6\times 10^{-4}\ \rm{M_{\odot}}$.
This amount of mass increases the spin period to $P_{\rm{spin}} \approx 683$ milliseconds if this mass is accreted from a neutrino cooled disc \citep{2015ApJ...798L..19M}.
During the second supernova, a small fraction of the fast ejecta ($\approx 10^{-4}- 10^{-3}\ \rm{M_{\odot}}$) is \textit{ballistically} accreted onto the pulsar companion, and therefore we do not expect it to recycle the pulsar.

\subsection{Conclusions}
Our understanding of merging binaries has come a long way since the discovery of gravitational waves almost 6 years ago, but these enigmatic sources continue to offer major puzzles and challenges. 
Our results suggest that ground-based facilities, like LIGO and Virgo, will 
detect these merging binary populations which have currently avoided detection in the Milky Way. Space- and ground-based observations over the coming decade should allow us to uncover the detailed nature of these most remarkable systems and provide us with an exciting opportunity to study novel regimes of binary stellar evolution.

\acknowledgments
The authors thank Ilya Mandel, Simon Stevenson, and the anonymous referee for useful discussions.
A.V.-G., S.L.S., and E.R.-R. acknowledge support by  Heising-Simons Foundation, the Danish National Research Foundation (DNRF132) and NSF (AST-1911206 and AST-1852393). 
D. R. A.-D. was supported by the Stavros Niarchos Foundation (SNF) and the Hellenic Foundation for Research and Innovation (H.F.R.I.) under the 2nd Call of ``Science and Society’' Action Always strive for excellence – ``Theodoros Papazoglou’' (Project Number: 01431).

%



\software{
Data and scripts used for this study available via  \url{zenodo.org/record/4682798}.
\textsc{GADGET-2} is publicly available at \url{https://wwwmpa.mpa-garching.mpg.de/gadget/}.
MESA is publicly available at \url{https://mesa.sourceforge.net}.
SPLASH is publicly available at \url{https://users.monash.edu.au/~dprice/splash/}.
}

\bibliographystyle{aasjournal}

\providecommand{\noopsort}[1]{}




\appendix

\section{1D evolution of stripped stars.}
\label{app:MESA}
We model the evolution of stripped stars using the 1D stellar evolution code MESA \citep{Paxton2011MESA} version 10398 \citep{Paxton2013MESA,Paxton2015MESA,Paxton2018MESA} as presented in \cite{AguileraDena2021}.
We follow the evolution from helium zero-age main sequence until the onset of core collapse, which we define as the moment where core infall velocity is larger than $1000\ \rm km\ s^{-1}$.

\subsection{Numerical setup.}
The initial models are created by artificially mixing hydrogen-rich models from the pre-main-sequence phase, and until the beginning of helium burning.
There is no mass loss until the beginning of helium burning, but the condition of homogeneity is relaxed at core nitrogen ignition; this guarantees the appropriate CNO element distribution (enhanced N, reduced C and O) for the stripped star.
We follow \cite{Yoon2017} to account for mass loss through stellar winds, dependent on the stellar type (WN or WC) and metallicity.
We use the \texttt{approx21} nuclear network and set resolution variables to \texttt{varcontrol\_target}=10$^{-5}$, and \texttt{mesh\_delta\_coef}=0.5, which results in a finer resolution than MESA's default.
Convection was modeled using standard mixing length theory \citep{1958ZA.....46..108B} with $\alpha_{MLT} = 2.0$, adopting the Ledoux criterion for instability, employing efficient semiconvection with $\alpha_{SC} = 1.0$ \citep{2019A&A...625A.132S}, and using predictive mixing in the helium burning regions \citep{Paxton2018MESA}.
We use MESA's \texttt{mlt++} for the treatment of energy transport in the envelope and neglect radiative acceleration in layers with $T>10^8$ K during late phases of evolution.
This results in compact helium zero-age main sequence radius of $\lesssim 1.2\ \rm{R_{\odot}}$, and a minimum mass threshold of 9.5 $\rm{M_{\odot}}$ for the $Z=0.02$ model.
We do not include convective overshooting, which could result in larger core masses for initially less massive stars.

\begin{figure}
    \centering
    \includegraphics[trim=0 7cm 0 7cm,clip,width=0.45\columnwidth]{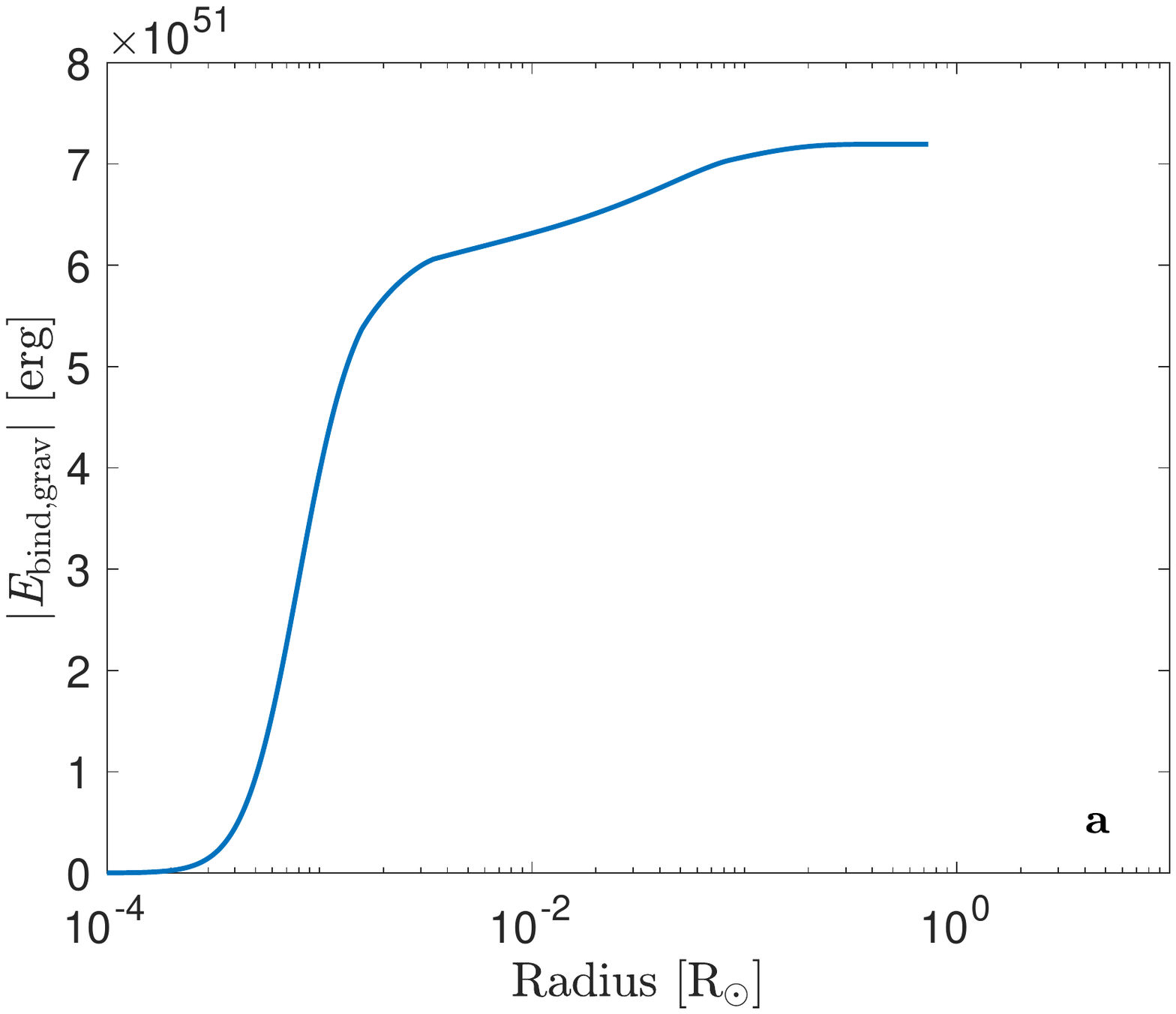}
    \includegraphics[trim=0 7cm 0 7cm,clip,width=0.45\columnwidth]{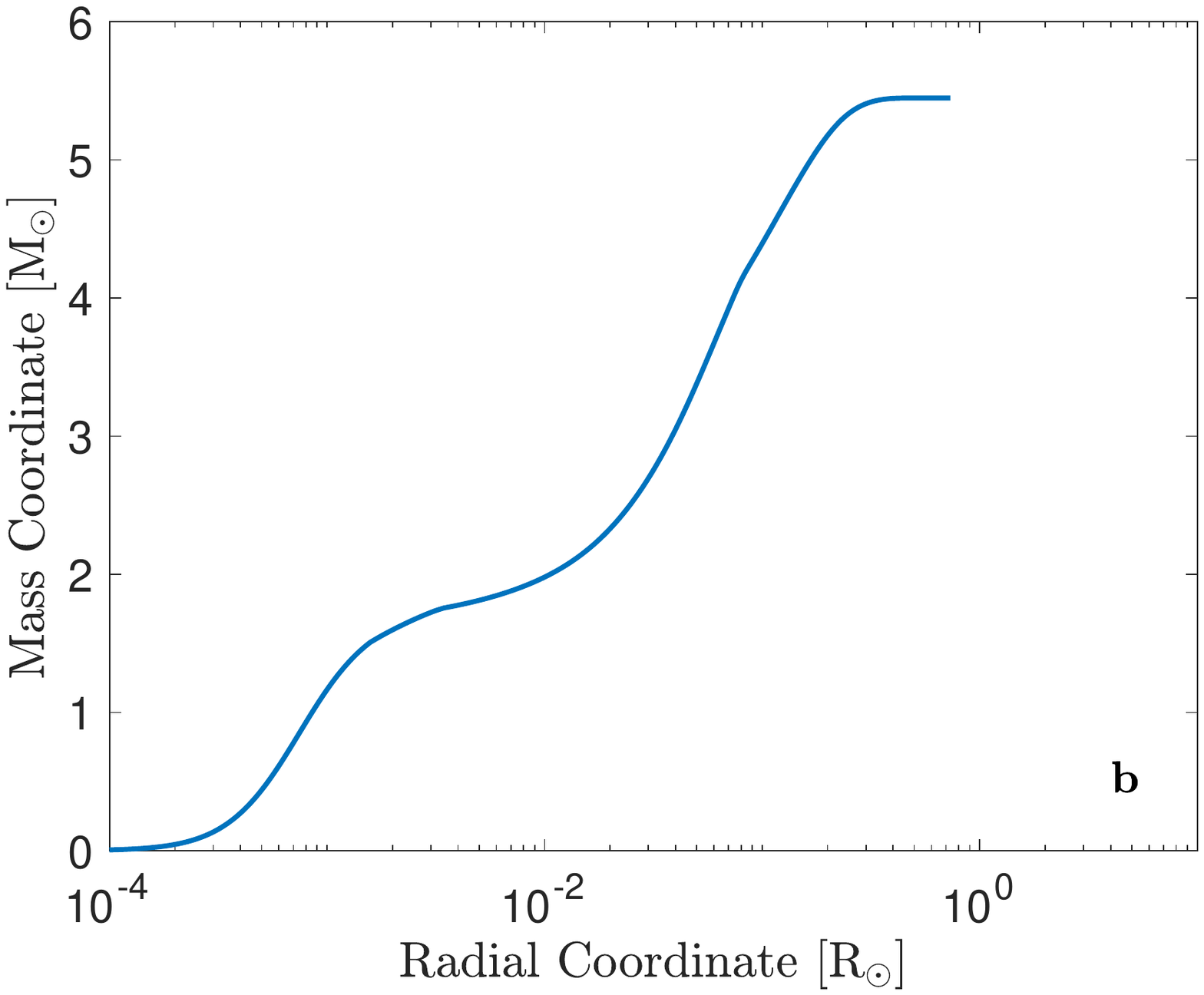}
    \includegraphics[trim=0 7cm 0 7cm,clip,width=0.45\columnwidth]{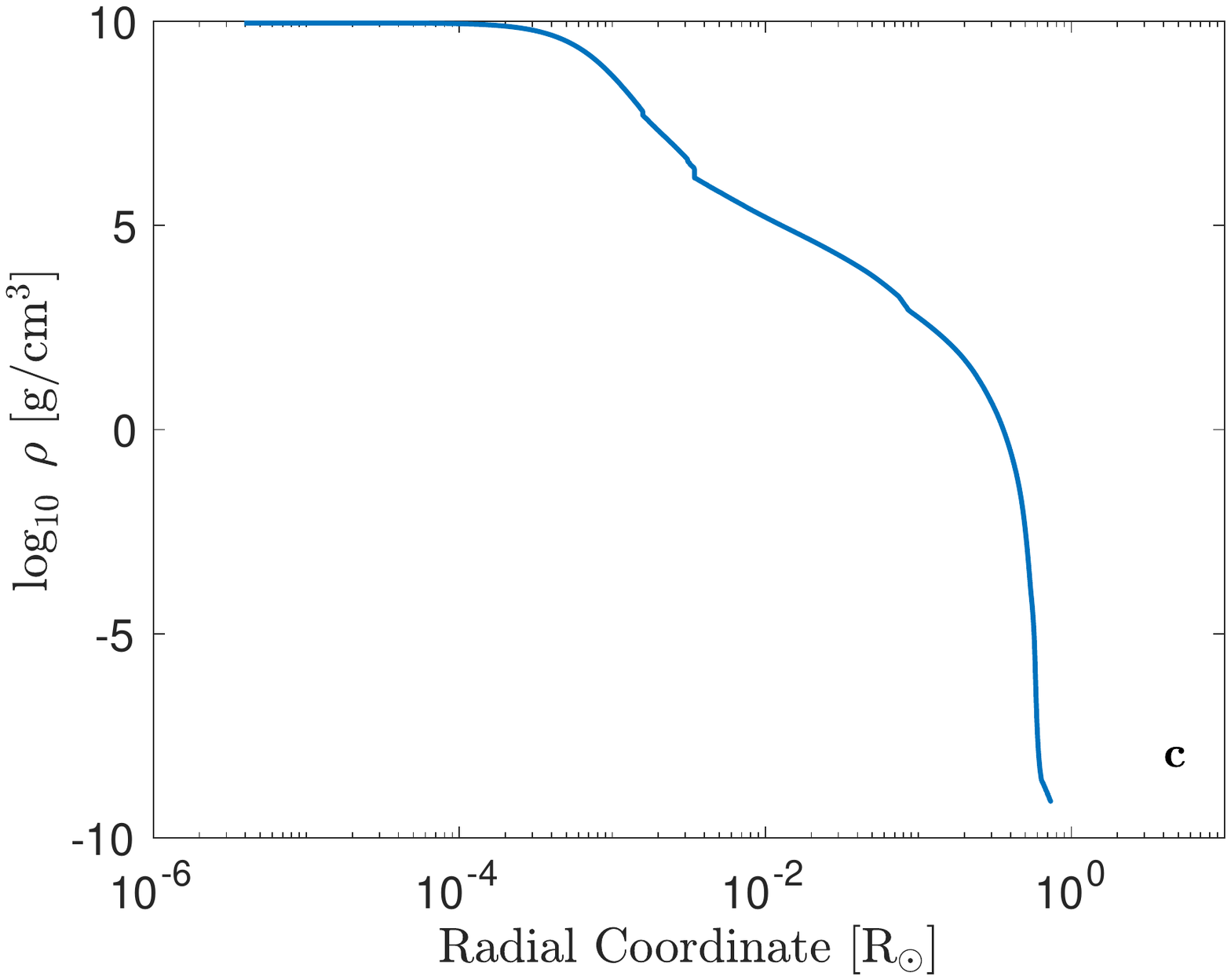}    
    \caption{
    Stellar structure of exploding model at the onset of core collapse.
    Gravitational binding energy (\textbf{a}), mass coordinate (\textbf{b}) and density (\textbf{c}) as a function of radial coordinate for the models with helium zero-age main sequence mass of 10.0 $\rm{M_{\odot}}$ at metallicity $Z=0.02$. 
    }
    \label{fig:app:stellar_structure}
\end{figure}

\subsection{Evolution of two representative models at $Z=0.02$.}
\begin{figure}
    \centering
    \includegraphics[trim=0 7cm 0 7cm,clip,width=\columnwidth]{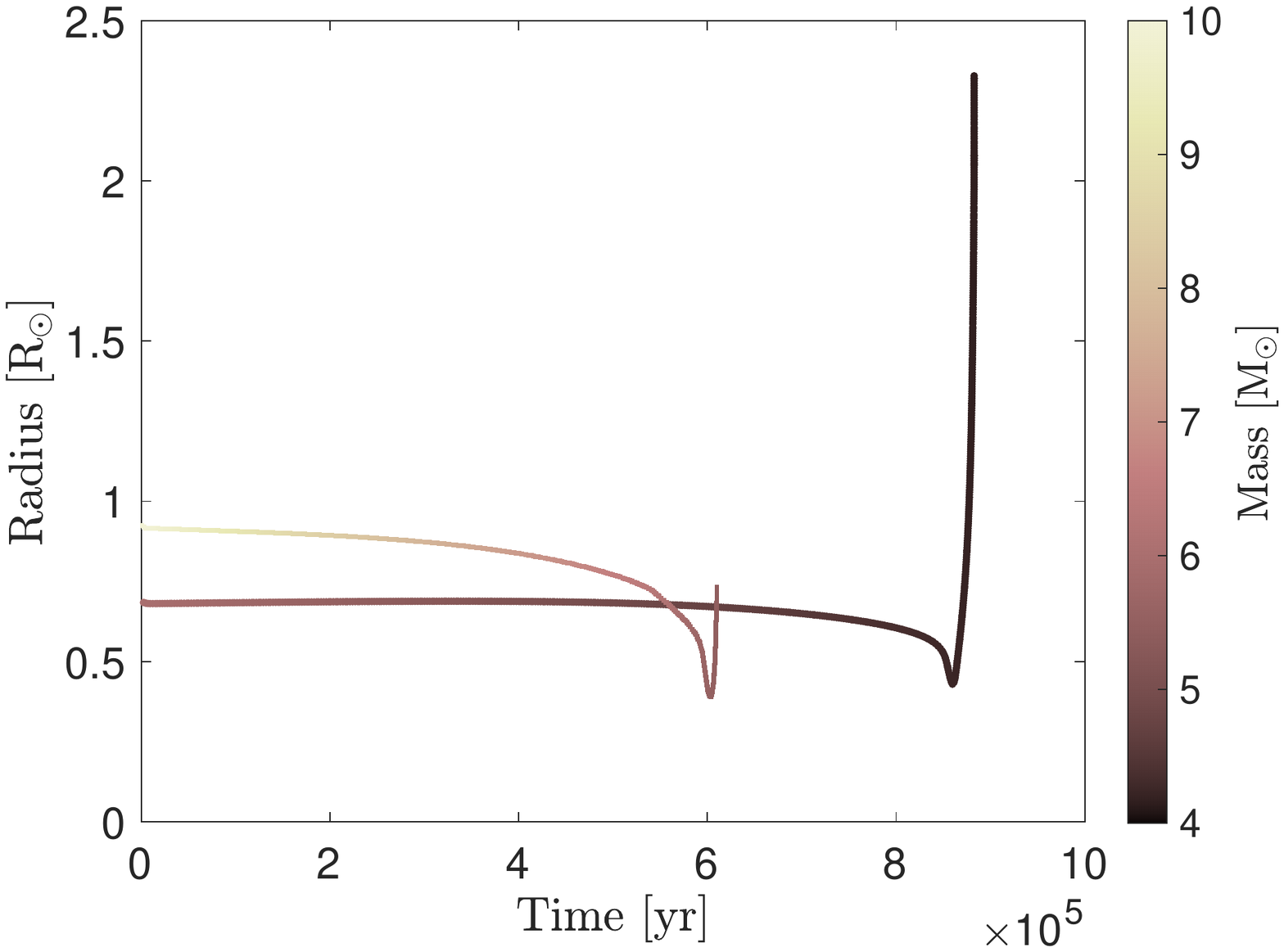}
    \caption{
    Time evolution of stripped stars.
    Radial (y-axis) and mass (colorbar) time evolution (x-axis) of two helium stars, from helium zero-age main sequence to core collapse, at metallicity $Z=0.02$.
    The initial helium star masses are 10.0 (initially more expanded) and 6.0 (initially more compact) $\rm M_{\odot}$, and reach core collapse with masses of 5.4 and 4.2 $\rm M_{\odot}$, respectively.
    The initially more expanded star contracts and the initially more compact star expands.
    }
    \label{fig:app:stripped_stars_time_evolution}
\end{figure}

The more massive model is initially 10.0 $\rm M_{\odot}$ and reaches advanced stages of burning faster and collapses before being able to expand above its initial radius (Figure~\ref{fig:app:stripped_stars_time_evolution}).
A 10.0 $\rm M_{\odot}$ helium core corresponds, for a single star, to a zero-age main sequence mass of $\approx 32.0\ \rm{M_{\odot}}$ \citep[according to the models from][]{Woosley2019}; however, the models presented here could have accreted matter via mass transfer episodes at some point in their lives.
At the end of the evolution, this model has a very compact envelope that decreases sharply in density until reaching the outer layers (Figure~\ref{fig:app:stellar_structure}).
The less massive model is 6.0 $\rm M_{\odot}$ and is computed to show the contrast with the more massive counterpart.
If this less massive model is in a close binary, it is likely to experience a mass transfer episode.
This less massive model is more similar to the canonical helium models that explain ultra-stripped stars, the progenitors of ultra-stripped supernovae, Galactic BNSs and GW170817 \citep{Tauris2013,Tauris2015ultra,Tauris2017formation,GW170817}.

\subsection{Metallicity and mixing study.}
\begin{figure}
    \centering
    \includegraphics[trim=0 7cm 0 7cm,clip,width=0.7\columnwidth]{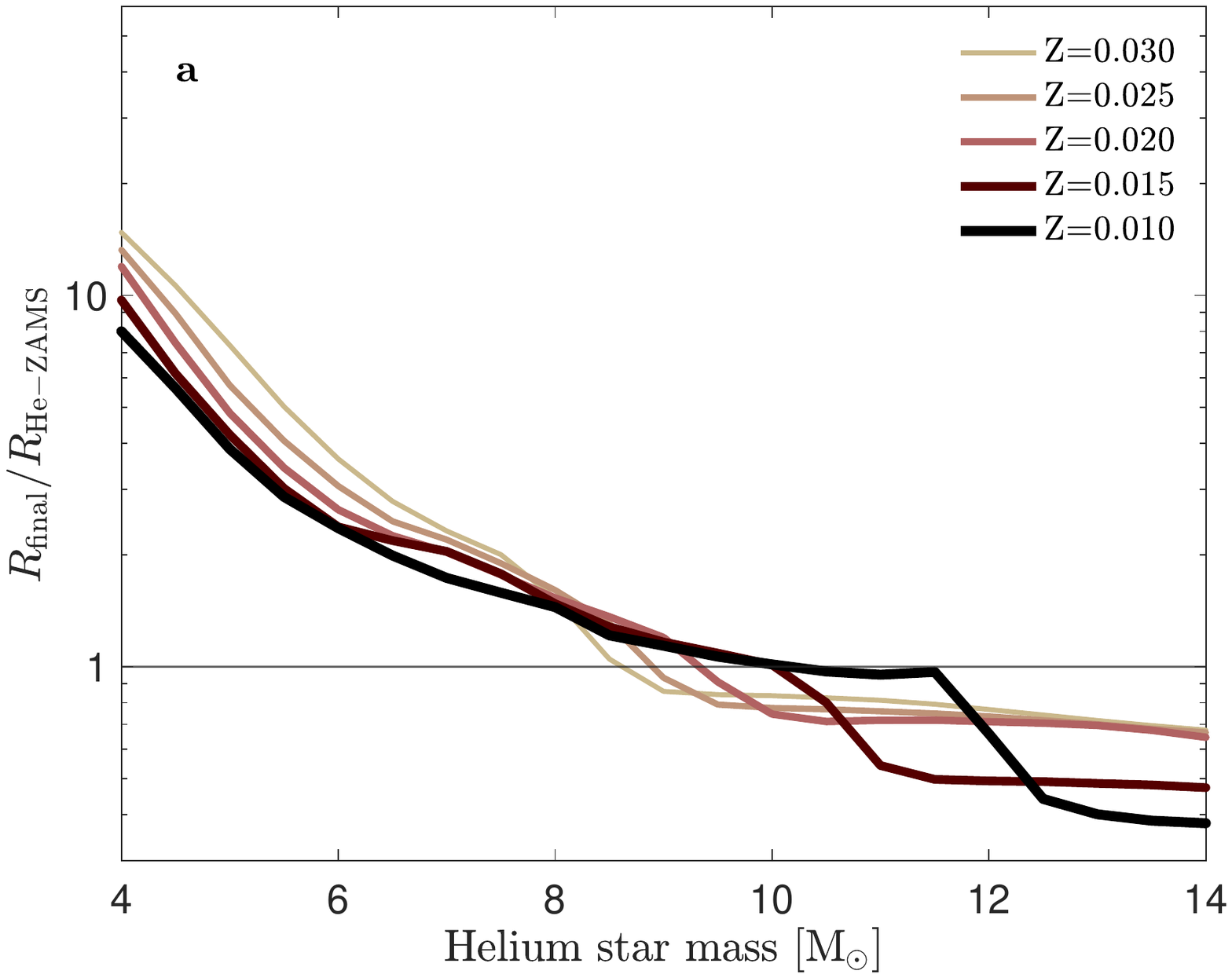}
    \includegraphics[trim=0 7cm 0 7cm,clip,width=0.7\columnwidth]{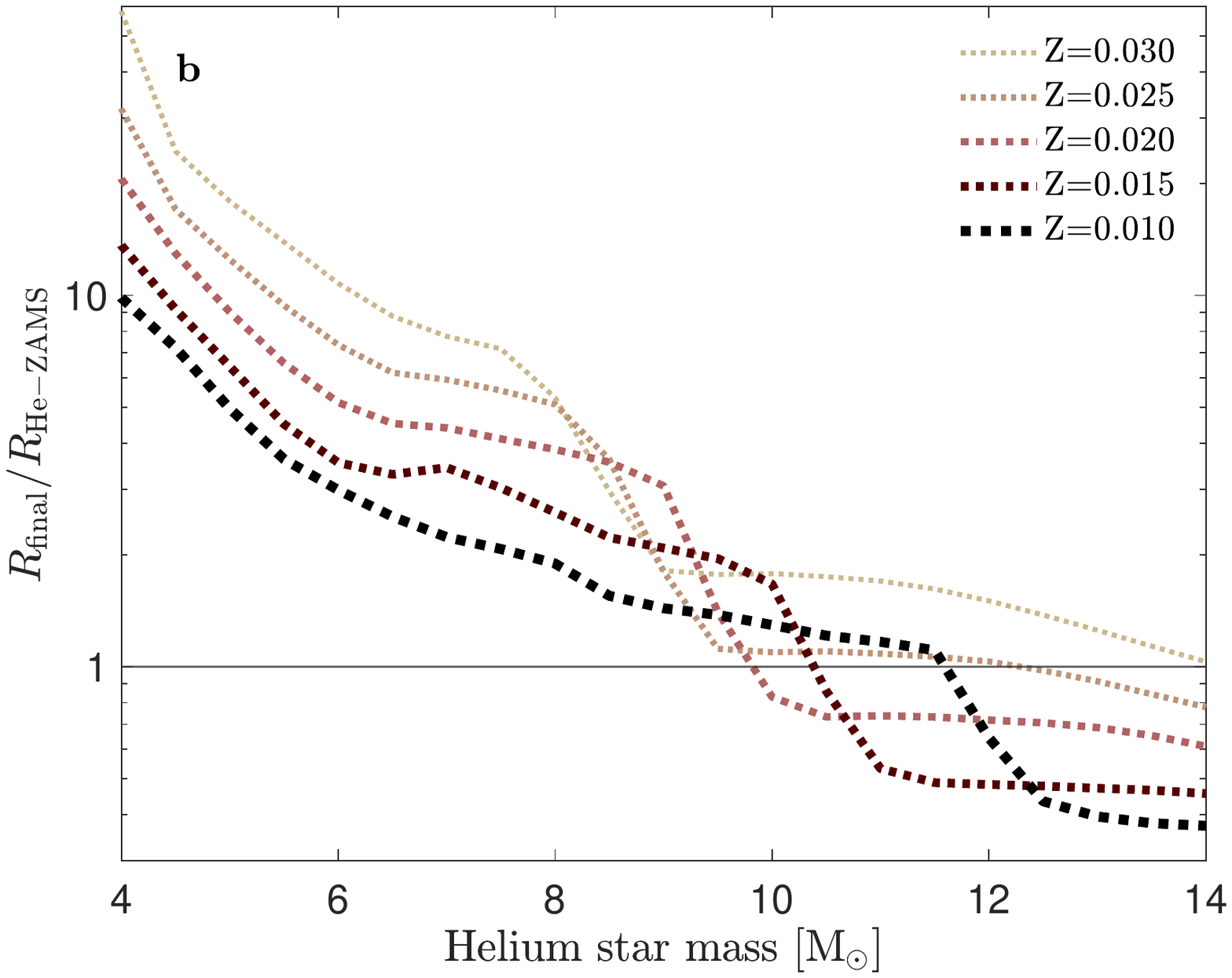}
    \caption{
    Summary of radial evolution of stripped helium stars.
    The behavior of the radial evolution of stripped stars is shown as a function of helium mass (x-axis) and metallicity (color).
    We parameterize the radii in terms of  $R_{\rm{He-ZAMS}}$ and $R_{\rm{final}}$ (see Methods).
    Stars remain compact when $R_{\rm{final}}/R_{\rm{He-ZAMS}}<1$; alternatively, significant expansion occurs when $R_{\rm{final}}/R_{\rm{He-ZAMS}}>1$.
    The results are shown in solid lines (\textbf{a}) for our standard model including \texttt{mlt++} and in dashed lines (\textbf{b}) for an alternative numerical treatment of mixing (Appendix~\ref{app:MESA}).
    The former ultimately leads to stars with less extended envelopes at lower masses.
    }
    \label{fig:app:mass_metallicity_stripped_stars}
\end{figure}

There is a dichotomy between stripped stars that do or do not expand which is mass, model and metallicity dependent \citep{Woosley2019}.
To test the mass and metallicity dependence we performed calculations for helium zero-age main sequence masses between $4.0 \leq M/\rm{M_{\odot}} \leq 14.0$ in steps of 0.5 $\rm{M_{\odot}}$ and at metallicities $Z=\{0.010,0.015,0.020,0.025,0.030\}$ (Figure~\ref{fig:app:mass_metallicity_stripped_stars}).
These are a subset of the simulations done in \cite{AguileraDena2021}.
Stripped stars have helium zero-age main sequence radii of $\lesssim 1.5\ \rm{R_{\odot}}$ and are more compact at lower metallicities.
In order to distinguish between stars which significantly expand and those which remain compact, we introduce a dimensionless factor $R_{\rm{final}}/R_{\rm{He-ZAMS}}$, where $R_{\rm{He-ZAMS}}$ is the radius at helium zero-age main sequence and $R_{\rm{final}}$ is the radius at the moment when the central carbon abundance is $\lesssim 5\times10^{-3}$, a proxy for central carbon depletion.
Stars with $R_{\rm{final}}/R_{\rm{He-ZAMS}}\lesssim 1$ remain compact.
The minimum mass threshold to remain compact is 9.0 $\rm{M_{\odot}}$ at $Z=0.02$.
We test for alternative energy transport envelope treatment by turning off \texttt{mlt++} and allowing for radiative acceleration in the envelope.
This variation results in helium zero-age main sequence radii of $\lesssim 1.5\ \rm{R_{\odot}}$, and mass threshold of 10.0 $\rm{M_{\odot}}$ for the $Z=0.02$ model.
The overall uncertainties on the minimum mass threshold are of order $\lesssim 1.0\ \rm{M_{\odot}}$.

\section{3D hydrodynamical simulation of fallback supernovae.}
\label{app:Gadget}
We study the explosion and fallback accretion of a stripped star with a neutron star companion using the 3D Lagrangian hydrodynamic SPH code \textsc{GADGET-2}  \citep{Springel2005Gadget}.
We use a modified version of \textsc{GADGET-2} that has been previously used to simulate supernovae in binary black hole forming binaries \citep{Batta2017,Schroder2018}.
Visualization of the hydrodynamical evolution (Figure \ref{fig:hydro}) was made using SPLASH \citep{Price2007}.

\subsection{Initial conditions and system properties.}
\begin{figure}
    \centering
    \includegraphics[trim=0 7cm 0 7cm,clip,width=0.75\columnwidth]{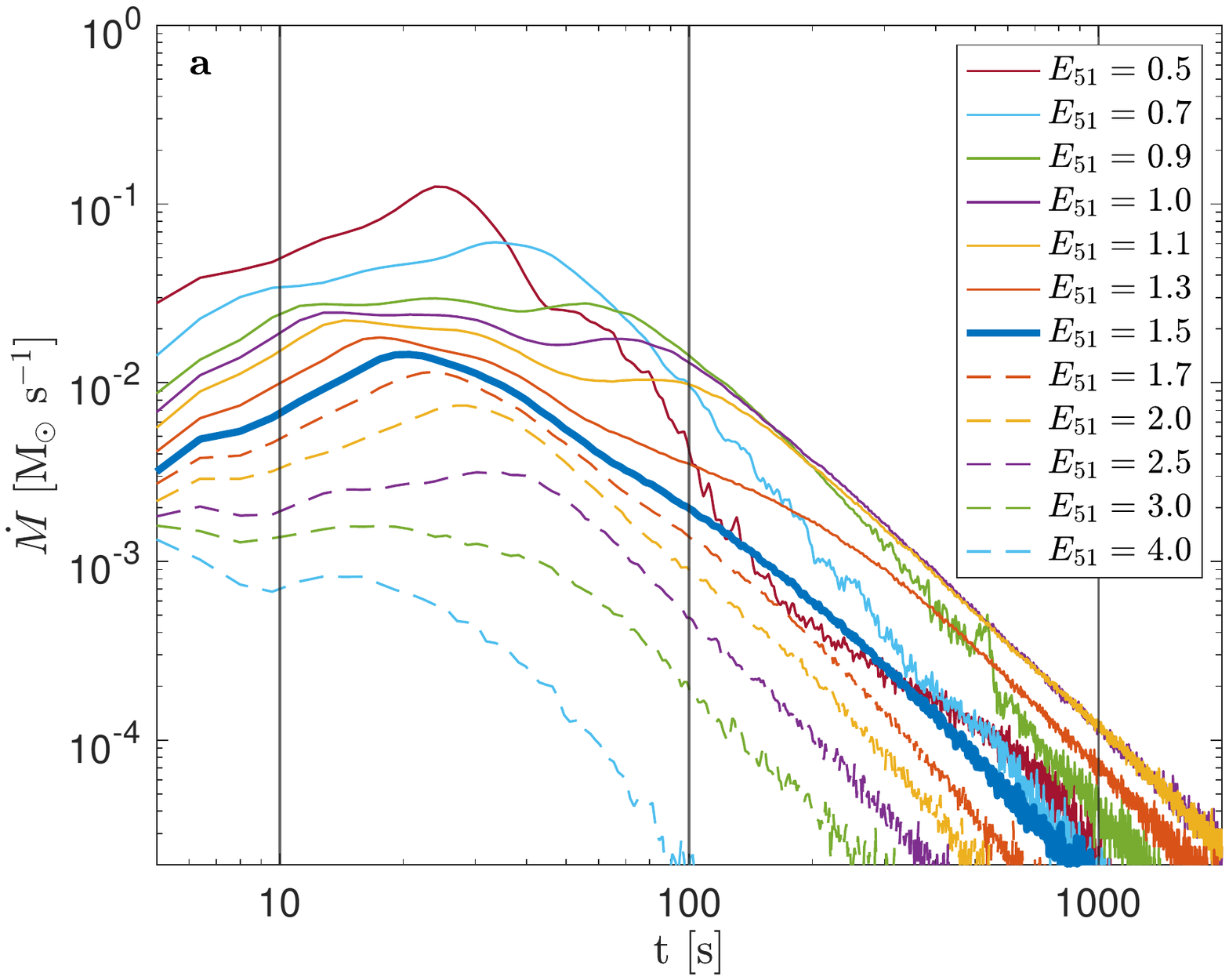}
    \includegraphics[trim=0 7cm 0 7cm,clip,width=0.75\columnwidth]{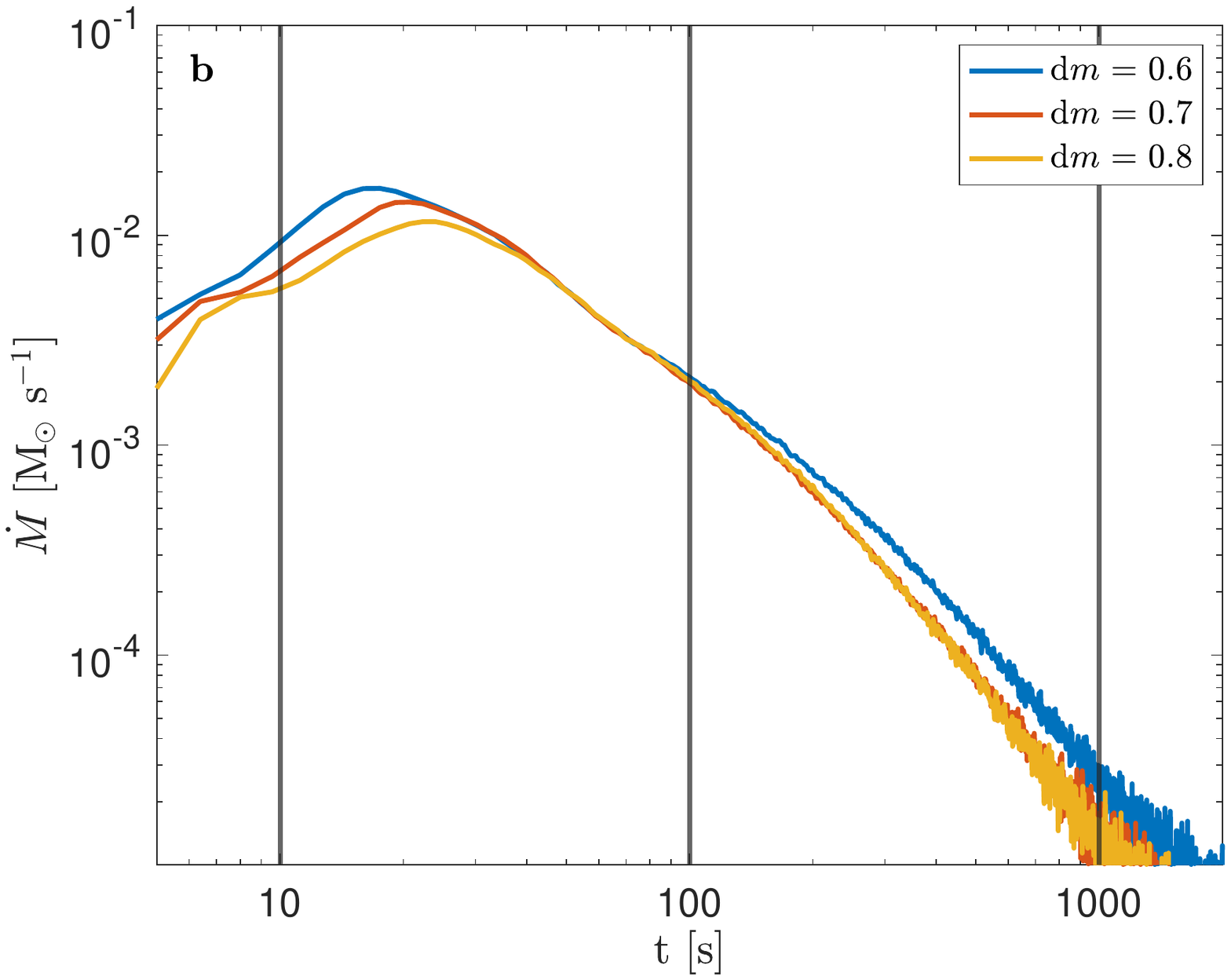}
    \caption{
    Fallback mass accretion rate of the exploding star.
    All models (\textbf{a}) and resolution study (\textbf{b}) exploring the evolution depending on the size of the mass shell where the kinetic explosion energy is deposited (Appendix~\ref{app:Gadget}). 
    }
    \label{fig:app:fallback}
\end{figure}

Here we describe the initial properties of our fiducial model.
The system is initialized as a circular gravitationally bound binary comprised of an exploding star and a neutron star companion at a separation of 1.4 $\rm{R_{\odot}}$.
The neutron star companion is defined as a sink particle type of mass $1.3\ \rm{M_{\odot}}$.
In order to build the initial conditions of the exploding star we use a 1D MESA model of a heavy compact progenitor at core collapse (Appendix~\ref{app:MESA}).
This progenitor, with a helium zero-age main sequence mass of 10.0 $\rm{M_{\odot}}$ and metallicity of $Z=0.02$, has mass of 5.4 $\rm{M_{\odot}}$ at core collapse.
The star’s final properties at core collapse are then mapped onto a 3D SPH particle distribution that reproduces the density profile. 
A million SPH particles are uniformly distributed on spherical shells generated with the HEALPix algorithm \citep{Gorski2005}. The shells are then spaced according to the local density \citep{Batta2017}. Due to the extremely low densities at the outer layers of the star, mapping with SPH particles became challenging. Therefore, we neglected low density material above $0.5\ \rm{R_{\odot}}$ resulting in $\approx 0.1\ \rm{M_{\odot}}$ artificially removed from the system (Figure~\ref{fig:app:stellar_structure}).
For the newly born neutron star, the innermost $1.3\ \rm{M_{\odot}}$ of the 3D stellar structure is removed and replaced by a sink particle with the same mass.
For our fiducial model (Figure~\ref{fig:hydro}) a kinetic explosion energy of 1.5 bethes is instantaneously deposited in the shell with mass $dm=0.7\ \rm{M_{\odot}}$ right above the $1.3\ \rm{M_{\odot}}$ that comprises the newly born neutron star.
We ran a series of models with different explosion energies (Figure~\ref{fig:explosions}) spanning from $0.5 \leq E_{\rm{exp}} \leq 4.0$ bethes resulting in different fallback evolution (Figure~\ref{fig:app:fallback}).

\subsection{Resolution study.}
\begin{figure}
    \centering
    \includegraphics[trim=0 7cm 0 7cm,clip,width=\columnwidth]{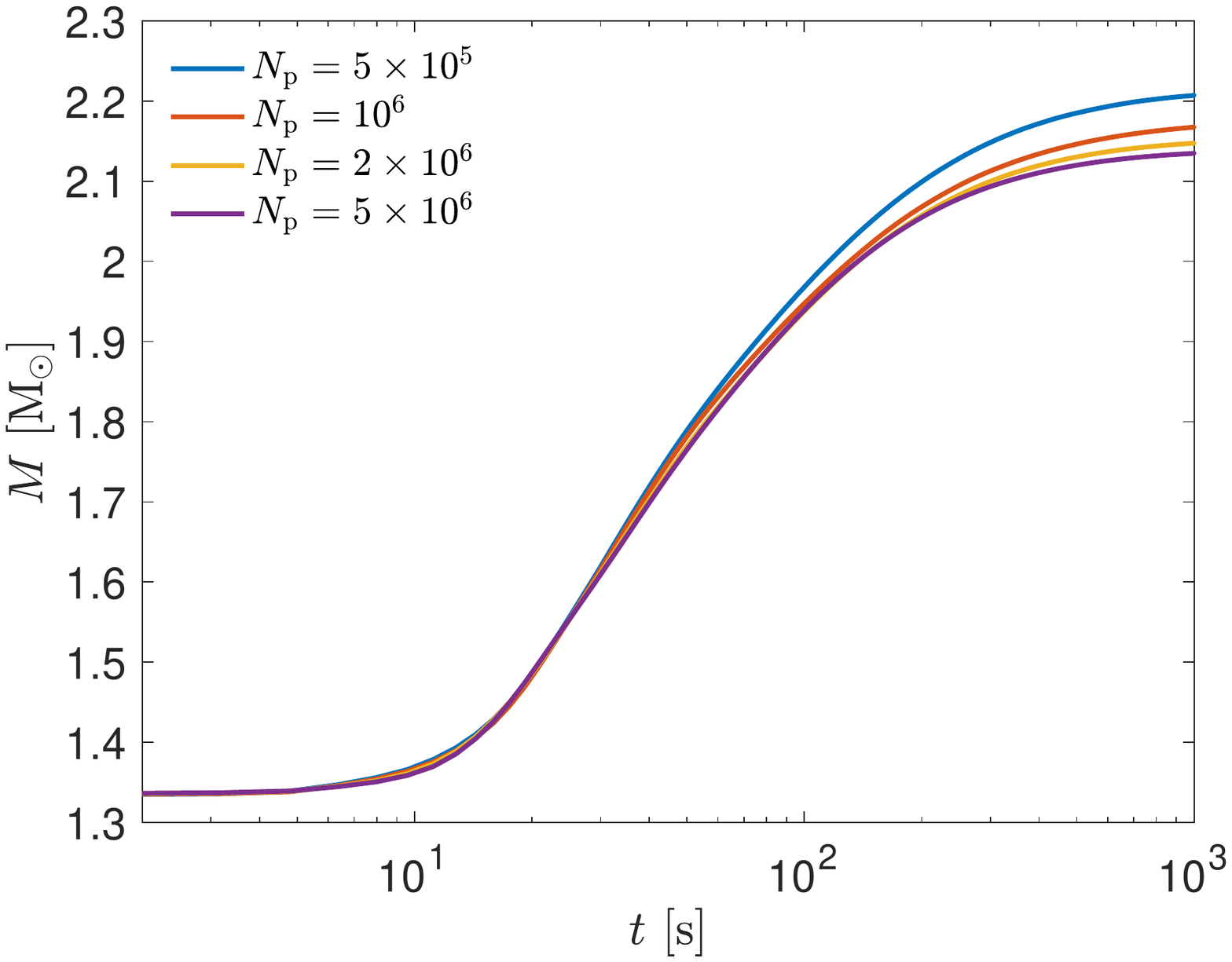}
    \caption{
    Post-supernova time evolution, including mass accretion, of the newly born neutron star.
    Number of particle ($N_{\rm{p}}$) resolution study to determine convergence in our simulations (Appendix~\ref{app:Gadget}).
    }
    \label{fig:app:convergence}
\end{figure}

We ran simulations for different resolutions to ensure that the remnant mass estimates are accurate for different choices of numerical parameters. 
For resolutions from $5\times10^5$ to $5\times10^6$ particles we found remnant mass variations smaller than $0.1\ \rm{M_{\odot}}$ and convergence as the number of particles increases (Figure~\ref{fig:app:convergence}). 
For our fiducial model we settled for a resolution of $10^6$ particles resulting in a mass difference of less than $0.04\ \rm{M_{\odot}}$ compared with the highest resolution.
The mass of shell in which the kinetic explosion energy is deposited is the main source of physical and numerical uncertainty. 
For the $E_{51}=1.5$ model, where $1\,E_{51}=1$ bethe, thin shell masses of $dm\approx 0.2\ \rm{M_{\odot}}$ lead to remnant masses of $\approx 3\ \rm{M_{\odot}}$, more than twice the remnant mass predicted by models which do not incorporate fallback \citep{Muller2016}.
Thicker shell masses of $dm\approx 0.7\ \rm{M_{\odot}}$ converge to more reasonable remnant masses of $\approx 2.1\ \rm{M_{\odot}}$ (Figure~\ref{fig:app:fallback}).
The mapping of 1D stellar models to 3D hydrodynamic ones is known to lead to discretization errors in the hydrostatic equilibrium \citep{Ohlmann2017}.
However, the effects of this mapping seem to be negligible in our simulations: while some of the outer layers of the star are artificially ejected because of this, the supernova of a non-exploding model is fully consistent with our lowest explosion energy model, implying complete fallback (see comment about equation on state in Section~\ref{sec:discAndConc}).
We lastly checked for any effect that a natal kick could have on the remnant mass.
Natal kicks of magnitudes of $\approx 10$, $\approx 100,$ and, $\rm \approx  1000\ km\ s^{-1}$ at random directions, which affect the orbit in timescales longer than the fallback timescale, made little difference with respect to our fiducial model. 

\subsection{Open questions in Supernova Explosion Mechanisms}

Supernovae are also very complicated processes to model numerically.
We do not present a self-consistent explosion model. Instead, we use a simplified approach to study the long-term evolution of supernova fallback in binaries and explore the sensitivity to the currently unknown supernova energy. 
This allows us to understand the role of fallback in creating light black holes rather than  heavy neutron star pairs.
These uncertainties in the explosion energy propagate directly into the rates estimates.
Moreover, because of the amount of ejected mass, it is more likely to have a binary remain bound the second explosion lead to a black hole instead of a neutron star.
Future observations will clarify the most likely outcome of stripped supernovae with neutron star companions and will allow us to place strict constraints on the explosion mechanism of massive stars.

Here we follow the model from \cite{Batta2017} in order to quantify the accretion history of the newly born neutron star.
We define an accretion radius $r_{\rm{acc}}<0.01\ \rm{R_{\odot}}$ from the edge of the innermost stable circular orbit (ISCO) of the compact object, in this case the $1.3\ \rm{M_{\odot}}$ proto-neutron-star.
Particles within the accretion radius and with less specific angular momentum ($j$) than the one needed to orbit ISCO are considered to be accreted, transferring their entire mass and angular momentum onto the compact object.
Particles within the accretion radius and with $j_{\rm{ISCO}} \leq j < 10\times j_{\rm{ISCO}}$ are assumed to be accreted via an accretion disc on a viscous timescale. To this end we neglect any additional feedback from this accretion.

\end{document}